\documentclass[preprint,showpacs,preprintnumbers,amsmath,amssymb]{revtex4}
\usepackage{graphics}
\usepackage{dcolumn}
\usepackage{bm}

\begin{document}

\title{Extension matrix representation theory of light beams and the Beauregard effect}

\author{Chun-Fang Li}

\affiliation{Department of Physics, Shanghai University, Shanghai 200444, P. R. China}
\affiliation{State Key Laboratory of Transient Optics and Photonics, Xi'an Institute of Optics and
Precision Mechanics of CAS, Xi'an 710119, P. R. China}


\begin{abstract}

It is shown that a light beam in free space is representable by an integral over a vectorial
angular spectrum that is expressed in terms of an extension matrix, which describes the
vectorial nature of the beam. A symmetry axis of the extension matrix is identified. When it
is neither perpendicular nor parallel to the propagation axis, we arrive at such beams that
show us for the first time the observable evidence of the Beauregard effect. The advanced
representation theory may yield any kinds of light beam, and the uncovered Beauregard effect
would play its unique roles in applications.

\end{abstract}

\pacs{42.25.Bs, 42.25.Ja, 02.10.Yn}
\maketitle


\paragraph*{Introduction}

The vectorial feature, or the polarization property, of a coherent light beam, which concerns
its angular momentum \cite{Allen-B, Santamato, Marrucci}, intensity distribution
\cite{Pattanayak, Davis-P, Dorn}, and diffraction characteristics \cite{Jordan, Greene}, has
become more and more important in diverse areas of applications, including optical tweezers
and spanner \cite{Simpson, Allen, Adachi}, optical data storage, optical trapping and
manipulation \cite{Ashkin, Xu}, and dark-field imaging \cite{Biss}. For a plane wave or a
nearly plane wave, we have the Jones vector \cite{Jones} to describe the vectorial feature of
the electric or magnetic field. But for a bound beam, we have not yet had such a clear concept
for its vectorial feature.

In this Letter, we advance a representation theory of free-space light beams in terms of an
extension matrix acting on a two-form angular spectrum. The extension matrix in this theory
plays the role of describing the vectorial feature of a bound beam. A symmetry axis of the
extension matrix is identified due to the transversality property of the electromagnetic wave
in free space. What is surprising is that when the symmetry axis of the extension matrix is
neither perpendicular nor parallel to the propagation axis of the beam, we arrive at such
beams that show us for the first time \cite{Li} the observable evidence of the Beauregard
effect. By Beauregard effect we mean a novel phenomenon, surmised more than 40 years ago by
Beauregard \cite{Beauregard}, that a circularly polarized beam of zero transverse wave vector
can be deflected in a transverse direction. It is shown that the transverse deflection due to
the Beauregard effect can be as large as the order of the beam waist under specific
conditions. The uncovered Beauregard effect would open a new area of applications.

\paragraph*{General theory}

We consider three-dimensional light beams that propagate in free space. Let us first write
out, in the Cartesian coordinate system, the integral expression for the vectorial electric
field of a beam that may propagate in positive $x$ direction,
\begin{equation}\label{electric field of a beam}
\mathbf{E}(\mathbf{x})= \int \int_{k^2_y+k^2_z<k^2} \frac{d k_y d k_z}{2\pi} \mathbf{A}
(k_y,k_z) \exp(i \mathbf{k} \cdot \mathbf{x}),
\end{equation}
where the time dependence $\exp(-i \omega t)$ is implied and omitted. As we know, any element
of the angular spectrum represented by its wave vector $
 \mathbf{k} \equiv (\begin{array}{ccc} k_x & k_y & k_z \end{array})^T
$ has only two independent polarization states, though its vectorial amplitude
$
 \mathbf{A} \equiv (\begin{array}{ccc} A_x & A_y & A_z \end{array})^T
$ has three components, where the superscript $T$ means transpose. Denoting respectively by
$s$ and $p$ the two orthogonal linear polarization states, we have $\mathbf{A}= \mathbf{A}_s+
\mathbf{A}_p= A_s \mathbf{s}+ A_p \mathbf{p}$, where $A_s$ and $A_p$ are the complex
amplitudes of the $s$ and $p$ polarization states, respectively, $\mathbf{s}$ and $\mathbf{p}$
are their respective unit vectors. Letting $\mathbf{s}=s_x \mathbf{e}_x+ s_y \mathbf{e}_y+ s_z
\mathbf{e}_z$ and $\mathbf{p}=p_x \mathbf{e}_x+ p_y \mathbf{e}_y+ p_z \mathbf{e}_z$, where
$\mathbf{e}_x$, $\mathbf{e}_y$, and $\mathbf{e}_z$ are the unit vectors in the directions of
the Cartesian coordinates, $s_j$ and $p_j$ ($j=x,y,z$) are real numbers, we have
\begin{equation} \label{vector in two-form}
\mathbf{A}= \mathrm{P} \tilde{A},
\end{equation}
where
\begin{equation}
\tilde{A}= \left(
                 \begin{array}{c}
                    A_s\\A_p
                 \end{array}
           \right)
\end{equation}
is the two-form amplitude \cite{Li} of the angular spectrum which is, or can be regarded as,
the Jones vector \cite{Jones} associated with the angular spectrum, and
\begin{equation} \label{extension matrix}
\mathrm{P}=\left(
                 \begin{array}{cc}
                     s_x & p_x\\s_y & p_y\\s_z & p_z
                 \end{array}
           \right)
          \equiv (
                  \begin{array}{cc}
                     \mathbf{s} & \mathbf{p}
                  \end{array}
                 )
\end{equation}
extends the two-form amplitude $\tilde{A}$ onto the three-component vectorial amplitude
$\mathbf{A}$ and is thus referred to as the extension matrix. It is clear that it is the
extension matrix rather than the two-form amplitude that describes the vectorial nature of the
beam. The mutual orthogonality between the unit vectors $\mathbf{s}$ and $\mathbf{p}$ and the
wave vector $\mathbf{k}$ leads to
\begin{eqnarray}
                  s^2_x+s^2_y+s^2_z & = & 1, \nonumber \\
                  p^2_x+p^2_y+p^2_z & = & 1, \nonumber \\
\label{relations} s_x p_x+s_y p_y+s_z p_z & = & 0,\\
                  k_x s_x+ k_y s_y+ k_z s_z & = & 0, \nonumber \\
                  k_x p_x+ k_y p_y+ k_z p_z & = & 0. \nonumber
\end{eqnarray}

We have represented a free-space light beam by Eqs. (\ref{electric field of a
beam})-(\ref{relations}). But there are only the above five equations to determine the six
real numbers $s_j$ and $p_j$ in the extension matrix. That is to say, there is one freedom to
be chosen. This can be done by specifying the symmetry axis of the extension matrix. Two
independent types of representation will be discussed below.

Without loss of generality, we consider the following two-form amplitude,
\begin{equation}\label{two-form amplitude}
\tilde{A}= \left( \begin{array}{c} l_1 \\ l_2  \end{array} \right) A \equiv \tilde{l} A,
\end{equation}
where $\tilde{l}= \left( \begin{array}{c} l_1 \\ l_2  \end{array} \right)$ describes the
polarization state of the angular spectrum and is assumed to satisfy the normalization
condition $\tilde{l}^{\dagger} \tilde{l}=1$, the superscript $\dagger$ means transpose
conjugate, and $A$ is the amplitude distribution of the angular spectrum (ADAS). For the sake
of simplicity, we will consider the following Gaussian distribution function \cite{Li},
\begin{equation}\label{angular distribution}
A=\left( \frac{w_y w_z}{\pi} \right)^{1/2}
    \exp\left[ -\frac{w^2_y}{2} (k_y-k_{y0})^2- \frac{w^2_z}{2} k^2_z\right],
\end{equation}
where $w_y=w_0/\cos \theta_0$, $w_z=w_0$, $w_0$ is half the width of the beam at waist, and
$k_{y0}=k \sin \theta_0$. This ADAS indicates that the axis of propagation, represented by its
principal wave vector $\mathbf{k}_0= (\begin{array}{ccc} k_{x0} & k_{y0} & 0
\end{array})^T$, is perpendicular to the $z$ axis, where $k_{x0}=k \cos \theta_0$. For the
same reason, only beams satisfying paraxial approximation \cite{Lax} $\Delta \theta=
\frac{1}{k w_0} \ll 1$ will be considered, where $\Delta \theta$ is half the divergence angle
of the beam. Under this condition, Eq. (\ref{electric field of a beam}) can be rewritten as
\cite{Enderlein}
\begin{equation}\label{electric field of paraxial beam}
\mathbf{E}(\mathbf{x})= \frac{1}{2\pi} \int_{-\infty}^{\infty} \int_{-\infty}^{\infty}
\mathbf{A} (k_y,k_z) \exp(i \mathbf{k} \cdot \mathbf{x})d k_y d k_z.
\end{equation}

\paragraph*{Representation of $p_z=0$ and uniformly polarized beams}

When $p_z=0$, we have for the other elements of the extension matrix that solve Eq.
(\ref{relations}),
\begin{eqnarray}
s_x & = & -\frac{k_x k_z}{k (k^2_x+ k^2_y)^{1/2}}, \nonumber \\
s_y & = & -\frac{k_y k_z}{k (k^2_x+ k^2_y)^{1/2}}, \nonumber \\
s_z & = & \frac{(k^2_x+ k^2_y)^{1/2}}{k}, \nonumber \\
p_x & = & -\frac{k_y}{(k^2_x+ k^2_y)^{1/2}}, \nonumber \\
p_y & = & \frac{k_x}{(k^2_x+ k^2_y)^{1/2}}. \nonumber
\end{eqnarray}
Denoting $\mathbf{k}_r \equiv k_r \mathbf{e}_r=k_x \mathbf{e}_x+ k_y \mathbf{e}_y$ in a
cylindrical coordinate system, where $k_x= k_r \cos \theta$, $k_y= k_r \sin \theta$, and
$\mathbf{e}_r$ is the unit vector in the radial direction, it is clear that $\mathbf{p}= p_x
\mathbf{e}_x+ p_y \mathbf{e}_y= \mathbf{e}_{\theta}$ is the unit vector in the azimuthal
direction. Furthermore, letting $\mathbf{s}_r= s_x \mathbf{e}_x+ s_y \mathbf{e}_y$, it is
found that $\mathbf{s}_r= -\frac{k_z}{k} \mathbf{e}_r$ is in the radial direction and that
$s_z=\frac{k_r}{k}$. All these facts show that the symmetry axis of the extension matrix
(\ref{extension matrix}) is the $z$ axis and thus is perpendicular to the axis of propagation.
Because the ADAS in Eq. (\ref{angular distribution}) is appreciable only in a small region in
which $|k_y-k_{y0}| \sim |k_z| \leq k \Delta \theta \ll k$, $k_z/k$ in the extension matrix
can be regarded as a small number in comparison with unity in view of the integral
(\ref{electric field of paraxial beam}). To the zeroth-order approximation, we have for the
extension matrix,
\begin{equation} \label{zeroth-order matrix for pz=0}
\mathrm{P} \approx \left(
                         \begin{array}{cc}
                            0 & -\sin \theta_0\\
                            0 & \cos \theta_0\\
                            1 & 0
                         \end{array}
                   \right).
\end{equation}
The electric field corresponding to extension matrix (\ref{zeroth-order matrix for pz=0}) is
\begin{eqnarray} \label{transverse electric field for pz=0}
\mathbf{E}_T (\mathbf{x}) & = & [l_1 \mathbf{e}_z+ l_2 (\mathbf{e}_y \cos \theta_0-
                                \mathbf{e}_x \sin \theta_0)] \nonumber \\
                          &   & \times \frac{1}{2\pi} \int \int A \exp(i \mathbf{k} \cdot \mathbf{x})d k_y d k_z,
\end{eqnarray}
which is transverse with respect to the propagation axis $\mathbf{k}_0$ and is uniformly
polarized. The polarization state is represented by that of the angular spectrum, $\tilde{l}$.
Higher-order corrections will be included if higher-order approximations are considered
\cite{Pattanayak}. When appropriate ADAS's are selected, the known kinds of uniformly
polarized beam will be obtained, including the fundamental Gaussian beam, Hermite-Gaussian
beams, Laguerre-Gaussian beams \cite{Kogelnik, Enderlein}, and the Bessel-Gaussian beams
\cite{Gori, Bagini}.

\paragraph*{Representation of $s_x=0$, non-uniformly polarized beams, and the Beauregard effect}

We have in this representation the following solution to Eq. (\ref{relations}),
\begin{eqnarray}
              s_y & = & -\frac{k_z}{(k^2_y+ k^2_z)^{1/2}}, \nonumber \\
              s_z & = & \frac{k_y}{(k^2_y+ k^2_z)^{1/2}}, \nonumber \\
\label{s_x=0} p_x & = & -\frac{(k^2_y+ k^2_z)^{1/2}}{k},  \\
              p_y & = & \frac{k_x k_y}{k (k^2_y+ k^2_z)^{1/2}}, \nonumber \\
              p_z & = & \frac{k_x k_z}{k (k^2_y+ k^2_z)^{1/2}}. \nonumber
\end{eqnarray}
Denoting $\mathbf{k}_r \equiv k_r \mathbf{e}_r = k_y \mathbf{e}_y+ k_z \mathbf{e}_z$ in
another cylindrical coordinate system, where $k_y=k_r \cos \varphi$, and $k_z=k_r \sin
\varphi$, we find that $\mathbf{s}= s_y \mathbf{e}_y+ s_z \mathbf{e}_z= \mathbf{e}_\varphi$ is
the unit vector in the azimuthal direction. In addition, letting $\mathbf{p}_r= p_y
\mathbf{e}_y+ p_z \mathbf{e}_z$, it is apparent that $\mathbf{p}_r= \frac{k_x}{k}
\mathbf{e}_r$ is in the radial direction and that $p_x=-\frac{k_r}{k}$. That is to say, the
extension matrix in this representation is symmetric with respect to the $x$ axis. Therefore,
$\theta_0$ here is the angle between the symmetry axis of the extension matrix and the
propagation axis. It will be shown in the following that different values of $\theta_0$
correspond to different beams with different polarization properties and intensity
distributions. All the beams in this representation are spatially nonuniformly polarized.

\subparagraph{The case of $\theta_0=0$ and axially symmetric polarization}

The symmetry axis of the extension matrix is parallel to the axis of propagation in this case.
We use the cylindrical coordinate system by defining $\mathbf{x}= x \mathbf{e}_x+ \mathbf{r}$,
where $\mathbf{r}= y \mathbf{e}_y+ z \mathbf{e}_z$, $y=r \cos \phi$, and $z= r \sin \phi$. The
ADAS in Eq. (\ref{angular distribution}) is expressed in the cylindrical coordinate system as
\begin{equation} \label{zero theta0}
A=\frac{w_0}{\sqrt{\pi}} \exp \left( -\frac{w^2_0}{2} k^2_r \right).
\end{equation}
Substituting Eqs. (\ref{vector in two-form}), (\ref{extension matrix}), (\ref{two-form
amplitude}), (\ref{s_x=0}), and (\ref{zero theta0}) into Eq. (\ref{electric field of paraxial
beam}) and with the help the following expansion,
\begin{equation} \label{expansion}
\exp(i \rho \cos \psi)= \sum_{m=-\infty}^{\infty} i^m J_m (\rho) \exp(im \psi),
\end{equation}
we obtain for the electric field
\begin{equation} \label{electric for zero theta0}
\mathbf{E}_1 (\mathbf{x})=  i (l_1 \mathbf{e}_{\phi}+ l_2 \mathbf{e}_r) r \chi_1 (r,x)
\exp(ikx)+  l_2 \mathbf{e}_x E^p_{1x} (\mathbf{x}),
\end{equation}
where
\begin{eqnarray}
\chi_1 (r,x) & = & \frac{1}{r} \int_0^{\infty} A_1 J_1 (r k_r) k_r dk_r= \frac{\sqrt{2} w_0}{4 w^3} \nonumber \\
             &   & \times \exp\left( -\frac{r^2}{4 w^2} \right)
                   \left[ I_0 \left( \frac{r^2}{4 w^2} \right)- I_1 \left( \frac{r^2}{4 w^2} \right) \right], \nonumber \\
E^p_{1x} (\mathbf{x}) & = & -\exp(ikx) \int_0^{\infty} \frac{k_r}{k} A_1 J_0 (r k_r) k_r dk_r, \nonumber \\
A_1 & = & A \exp \left( -\frac{ix}{2k} k^2_r \right), \nonumber \\
w(x) & = & w_0 \left( 1+i \frac{x}{k w^2_0} \right)^{1/2}, \nonumber
\end{eqnarray}
$J_m$'s are the Bessel functions of the first kind, $I_0$ and $I_1$ are the modified Bessel
functions of the first kind. In deriving Eq. (\ref{electric for zero theta0}), we have made
(i) the paraxial approximation \cite{Enderlein} $k_x \approx k- \frac{k^2_r}{2k}$ in the
exponential factor $\exp(i \mathbf{k} \cdot \mathbf{x})$, (ii) and the zeroth-order
approximation $\mathbf{p}_r= \frac{k_x}{k} \mathbf{e}_r \approx \mathbf{e}_r$ in the extension
matrix.

Eq. (\ref{electric for zero theta0}) describes beams of axially symmetric polarization. The
first term on the right side is the transverse component and is of the zeroth order. The
second term is the longitudinal component. It is of the first order, $\sim k_r/k$, and is thus
much smaller than the transverse component \cite{Lax, Davis}. Neglecting the small
longitudinal component, the beam is dark on the principal axis $r=0$. Locally, it is
elliptically polarized the same as the angular spectrum. Other kinds of axially symmetrically
polarized beams will be obtained when appropriate ADAS's are selected, including the
azimuthally and radially polarized Bessel-Gaussian and Laguerre-Gaussian beams \cite{Hall,
Tovar, Kozawa}.

\subparagraph{The case of $|\theta_0| \leq \Delta \theta$ and the Beauregard effect}

If $|\theta_0|$ is not equal to zero but is very small satisfying $|\theta_0| \leq \Delta
\theta$, we have approximately $w_y \approx w_0$, so that the ADAS in Eq. (\ref{angular
distribution}) becomes
\begin{equation} \label{small theta0}
A=A_0 \exp(k_{y0} w^2_0 k_y),
\end{equation}
where
$$
A_0= \frac{w_0}{\sqrt{\pi}} \exp\left( -\frac{w^2_0}{2} k^2_{y0} \right) \exp\left(
-\frac{w^2_0}{2} k^2_r \right).
$$
With the help of expansion (\ref{expansion}) and the Graf formula \cite{Andrews}, we obtain
for the beam's electric field after substituting Eqs. (\ref{vector in two-form}),
(\ref{extension matrix}), (\ref{two-form amplitude}), (\ref{s_x=0}), and (\ref{small theta0})
into Eq. (\ref{electric field of paraxial beam}),
\begin{equation} \label{electric for small theta0}
\mathbf{E}_2 (\mathbf{x})= \mathbf{E}_{2T} (\mathbf{x})+ l_2 \mathbf{e}_x E^p_{2x}
(\mathbf{x}),
\end{equation}
where
\begin{eqnarray}
\mathbf{E}_{2T} (\mathbf{x}) & = & [i(l_1 \mathbf{e}_{\phi}+ l_2 \mathbf{e}_{r}) r+ (l_1
\mathbf{e}_z+ l_2 \mathbf{e}_y) k_{y0} w^2_0] \nonumber \\
                             &   & \times \chi_2 (\gamma,x) \exp(ikx), \nonumber \\
E^p_{2x} (\mathbf{x}) & = & -\exp(ikx) \int_0^{\infty} \frac{k_r}{k} A_2 J_0 (\gamma k_r) k_r dk_r, \nonumber \\
\chi_2 (\gamma,x) & = & \frac{1}{\gamma} \int_0^{\infty} A_2 J_1 (\gamma k_r) k_r dk_r \nonumber \\
                  & = & \frac{\sqrt{2} w_0}{4 w^3} \exp \left( -\frac{w^2_0}{2} k^2_{y0} \right)
                        \exp \left( -\frac{\gamma^2}{4 w^2} \right) \nonumber \\
                  &   & \times \left[ I_0 \left( \frac{\gamma^2}{4 w^2} \right)-
                        I_1 \left(\frac{\gamma^2}{4 w^2} \right) \right], \nonumber \\
A_2 & = & A_0 \exp \left( -\frac{ix}{2k} k^2_r \right), \nonumber \\
\gamma & = & (r^2- k^2_{y0} w^4_0-2i k_{y0} w^2_0 r \cos \phi)^{1/2}. \nonumber
\end{eqnarray}
In deriving Eq. (\ref{electric for small theta0}), we have also made the aforementioned two
approximations.

The first term on the right side of Eq. (\ref{electric for small theta0}) is the transverse
component, and the second term is the longitudinal component, which is also of the first order
as before. It is interesting to note that the transverse component consists of two parts
having different polarization symmetries. One is of axially symmetric polarization and is dark
on the propagation axis $r=0$. The other is uniformly polarized. But locally both of them have
the same polarization state as that of the angular spectrum, $\tilde{l}$. It is the
interference between those two parts that yields the Beauregard effect as will be shown below.

Neglecting the small longitudinal component, the intensity distribution of the beam is found
to be
\begin{equation} \label{intensity distribution formula}
I_2 \approx |\mathbf{E}_{2T}|^2= (y^2+z^2+2 \sigma k_{y0} w^2_0 z + k^2_{y0} w^4_0)|\chi_2|^2,
\end{equation}
where $\sigma= i (l^*_1 l_2-l_1 l^*_2)$ is the polarization ellipticity of the angular
spectrum. The first factor on the right side of Eq. (\ref{intensity distribution formula})
shows that the intensity is minimum at a point $\mathbf{r}_0= -\mathbf{e}_z \sigma k_{y0}
w^2_0$ on the $z$ axis. The displacement of this point from the propagation axis is
\begin{equation} \label{depression position}
z_0= -\sigma k_{y0} w^2_0= - \sigma k w^2_0 \sin \theta_0.
\end{equation}
For angular spectra of circular polarizations $\sigma= \pm 1$, the minimum intensity is equal
to zero. The single-sided depression of the beam intensity on the $z$ axis renders the beam
centroid deflected transversely in the $z$ direction unless $\sigma$ or $\theta_0$ is equal to
zero. According to Eq. (\ref{depression position}), the deflection will be flipped to the
opposite side of the propagation axis by changing either the sign of $\sigma$ or the sign of
$\theta_0$. Since the expectation value of the transverse wave vector $k_z$ of the beam
vanishes, the transverse deflection of the beam centroid in the $z$ direction is nothing but
the Beauregard effect \cite{Li}. To the best of my knowledge, this is the first time to
demonstrate the observable evidence of the Beauregard effect.
\begin{figure}[ht]
\scalebox{0.7}[0.7]{\includegraphics{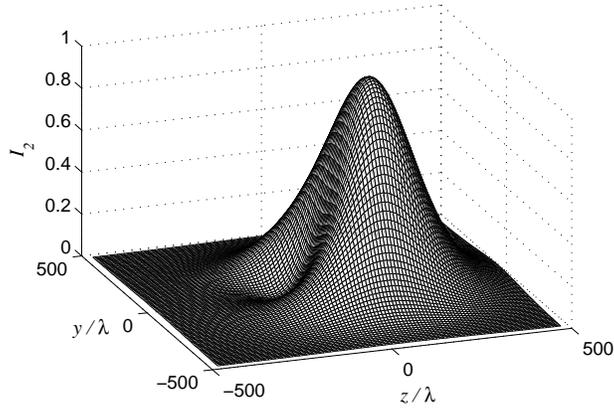}} \caption{Normlized intensity distribution on
plane $x=0$ for $\sigma=1$ in representation $s_x=0$, where $\theta_0= \Delta\theta= 10^{-3}$
rad, the $y$ and $z$ coordinates are in units of the wavelength $\lambda$.} \label{intensity
distribution figure}
\end{figure}
As an example, we show in Fig. \ref{intensity distribution figure} the normalized intensity
distribution of the beam on plane $x=0$, where $\sigma=1$, and $\theta_0= \Delta \theta=
10^{-3}$ rad. The intensity in Fig. \ref{intensity distribution figure} is numerically
calculated directly from Eq. (\ref{electric for small theta0}) and is well approximated by Eq.
(\ref{intensity distribution formula}) as numerical calculations show. The null point is
located at $z_0 \approx -w_0= -159 \lambda$ on the $z$ axis. As a result, the beam centroid is
deflected in the positive $z$ direction by $\langle z \rangle \approx 102 \lambda$ as can be
obtained by using formula (24) of Ref. \cite{Li}, which is as large as the order of $w_0$.

\subparagraph{The case of $|\theta_0| \geq \Delta \theta$}

The electric field distribution of the beam in this case cannot be approximately expressed so
clearly as have been done above. But the propagation characteristics of a paraxial beam can be
studied numerically by use of the integral (\ref{electric field of paraxial beam}). For a
given $\theta_0$, the beam deflection in the transverse direction due to the Beauregard effect
can be calculated by using formula (24) of Ref. \cite{Li}.

In conclusion, the representation theory advanced here may produce any kinds of free-space
light beam that solve the Maxwell equations. The Beauregard effect uncovered here would be
closely related to the Imbert-Fedorov effect \cite{Li} and open a new area of applications. It
also has vital implications to the counterpart of the quantum-mechanical matter waves
\cite{Beauregard2}. In order to explore the Beauregard effect in experiments, one should be
able to control separately the propagation axis of the beam, the symmetry axis of the
extension matrix, as well as the polarization ellipticity of the angular spectrum.

This work was supported in part by the National Natural Science Foundation of China
(60377025), Science and Technology Commission of Shanghai Municipal (04JC14036), and the
Shanghai Leading Academic Discipline Program (T0104).

\end{document}